\documentclass{emulateapj}
\usepackage{graphicx}

\shorttitle{Anisotropy vs chemical composition} 
\shortauthors{Lemoine \& Waxman}

\begin{document}


\title{Anisotropy vs chemical composition at ultra-high energies}


\author{Martin Lemoine\altaffilmark{1}}
\affil{Institut d'Astrophysique de Paris, \\CNRS, Universit\'e Pierre \& Marie Curie,\\
98 bis boulevard Arago, F-75014 Paris, France}
\altaffiltext{1}{email: lemoine@iap.fr}
\author{Eli Waxman\altaffilmark{2}} \affil{Physics Faculty,\\ Weizmann
  Institute,\\ Rehovot 7600, Israel} \altaffiltext{2}{email:
  eli.waxman@weizmann.ac.il}




\begin{abstract}
This paper proposes and discusses a test of the chemical composition
of ultra-high energy cosmic rays that relies on the anisotropy
patterns measured as a function of energy. In particular, we show that
if one records an anisotropy signal produced by heavy nuclei of charge
$Z$ above an energy $E_{\rm thr}$, one should record an even stronger
(possibly much stronger) anisotropy at energies $>E_{\rm thr}/Z$ due
to the proton component that is expected to be associated with the
sources of the heavy nuclei. This conclusion remains robust with
respect to the parameters characterizing the sources and it does not
depend at all on the modelling of astrophysical magnetic fields. As a
concrete example, we apply this test to the most recent data of the
Pierre Auger Observatory. Assuming that the anisotropy reported above
55~EeV is not a statistical accident, and that no significant
anisotropy has been observed at energies $\lesssim 10\,$EeV, we show
that the apparent clustering toward Cen~A cannot be attributed to
heavy nuclei. Similar conclusions are drawn regarding the apparent
excess correlation with nearby active galactic nuclei. We then discuss
a robust lower bound to the magnetic luminosity that a source must
possess in order to be able to accelerate particles of charge $Z$ up
to $100\,$EeV, $L_B\,\gtrsim\,10^{45}\,Z^{-2}\,$erg/s. Using this
bound in conjunction with the above conclusions, we argue that the
current PAO data does not support the model of cosmic ray origin in
active radio-quiet or even radio-loud galaxies. Finally, we
demonstrate that the apparent clustering in the direction of Cen~A can
be explained by the contribution of the last few gamma-ray bursts or
magnetars in the host galaxy thanks to the scattering of the cosmic
rays on the magnetized lobes.

\end{abstract}




\section{Introduction}
The sources of ultra-high energy cosmic rays have remained elusive in
spite of the enormous progress reached on the experimental side, with
present day detectors reaching apertures
$>\,10\,000\,$km$^2$~sr~yr. The differential energy spectrum, the
chemical composition and the distribution of arrival directions on the
sky are as many clues to the nature of the source. There is now a
consensus on the existence of the GZK cut-off (Greisen 1966; Zatsepin
\& Kuz'min 1966), which has been observed by two different experiments
(Abbasi et al. 2008, Abraham et al. 2008a). The existence of this GZK
cut-off puts on solid ground the models which attribute the origin of
ultra-high energy cosmic rays to powerful astrophysical objects
distributed on cosmological scales, such as powerful radio-galaxies
(Rachen \& Biermann 1993), gamma-ray bursts (Milgrom \& Usov 1995,
Vietri 1995, Waxman 1995), or magnetars (Arons 2003).

Experimental results on the chemical composition and anisotropies are
however far more confusing. While the most recent analysis of HiRes
data points to a proton composition above the ankle (Abbasi et
al. 2005, Belz 2009), the fluorescence data of the Pierre Auger
Observatory (PAO) rather indicates an increasingly heavier composition
above this energy, with the last data points at $\sim\,30-50\,$EeV
close to expectations for iron (Unger et al. 2007, Wahlberg et
al. 2009). Regarding the distribution of arrival directions on the
sky, there exist various contradictory claims, see for instance the
correlation with the super-galactic plane reported by Stanev et
al. (1995), contradicted by the results of the AGASA experiment
(Takeda et al. 1999); the claim for a correlation with BL Lac objects
in Tinyakov \& Tkachev (2001, 2002), Gorbunov et al. (2002, 2004),
which has been debated (Evans, Ferrer \& Sarkar 2003, 2004; Tinyakov
\& Tkachev 2004); as well as the possible detections of multiplets in
various datasets, e.g. Takeda et al. (1999), Uchihori et al. (2000),
Farrar, Berlind \& Hogg (2006), the significance of which is
questioned in Abbasi et al. (2004) and Finley \& Westerhoff (2004);
or, finally, the recent data of the PAO, which reveal a correlation
with nearby active galactic nuclei (AGN), the maximal signal being
obtained for a search radius of $3.1^\circ$ around AGN located closer
than 75~Mpc and for energies above $57\,$EeV (Abraham et al. 2007,
2008b).

In general, these issues of anisotropies and chemical composition are
discussed separately. However, as we argue in the present paper, one
can construct a powerful test of the chemical composition by using the
anisotropy signal at various energies. This becomes particularly
interesting when one notes that measurements of chemical composition
are relatively sensitive to the details of the shower reconstruction,
in particular to the extrapolation of hadronic models at energies
beyond those currently tested in accelerators (see Wibig 2008, Ulrich
et al. 2009a,b and Wibig 2009 for recent discussions of this
issue). Moreover, measurements of chemical composition rely on the use
of fluorescence detectors whose duty cycle is rather low and cannot be
made event by event, so that they are limited by statistics.  

The essence of the test proposed in this paper is to exploit the fact
that a source in the sky emitting heavy nuclei of charge $Z$ at an
energy $E$ is expected to produce a similar anisotropy pattern at
energies $E/Z$ via the proton component which is expected to be
associated with the same source. Our expectation for the existence of
a proton component relies on the theoretical expectation, that if
protons are present in the plasma in which particle acceleration takes
place, then they should be accelerated along side with the heavy
nuclei.  As discussed in some detail at the end of
\S~\ref{sec:test-chem}, a source accelerating heavy nuclei to energy
$E$ is quite generally expected to accelerate protons to energy $E/Z$.
Although the isotropic background "noise" increases at lower energies,
we show in \S~\ref{sec:test-chem} that the signal-to-noise ratio for
the detection of the anisotropy also increases. This prediction does
not depend at all on the modelling of astrophysical magnetic fields as
it only relies on the property that protons of energy $E/Z$ follow the
same path in the intervening magnetic fields and produce the same
angular image as heavy nuclei of charge $Z$ and energy $E$. The
proposed test is discussed in more detail in
Section~\ref{sec:test-chem}. When applied to the most recent data of
the PAO (Section~\ref{sec:pao}), it shows that the signal that is
responsible for the apparent anisotropy pattern at energies $>55\,$EeV
must not be heavy but light nuclei, provided this anisotropy is
confirmed by future data, and provided the PAO does not see evidence
for anisotropy at lower energies, as seems to be the case (Abraham et
al. 2007).

We discuss the implications of these results in the last two
Sections. In Section~\ref{sec:agn}, we argue that local AGN (including
FR~I radio-galaxies) do not possess the power required to accelerate
protons to ultra-high energies. When combined with the previous
conclusions, this leads to the conclusion that the current PAO data do
not support AGN as sources of the highest energy cosmic rays. Finally,
Section~\ref{sec:disc} summarizes the findings and concludes that the
current data on composition and anisotropy suggest one of the
following: (i) the shower modelling or, what would be more interesting,
the hadronic theoretical models of shower reconstruction are in error
at high energy; (ii) the composition switches abruptly from heavy to
light above $30-50\,$EeV; (iii) the source injects primarily heavy
nuclei (which seems unlikely); (iv) the anisotropy seen by the PAO is
a statistical artefact.
 
\section{Testing the chemical composition of ultra-high energy cosmic
  rays with anisotropy data}\label{sec:test-chem}

As argued below, one can use the results of searches for anisotropy at
various energies in order to constrain the chemical composition of
ultra-high energy cosmic rays. The basic claim is the following: if
one detects anisotropy above some energy $E_{\rm thr}$, but not below,
then the chemical composition is most likely light above $E_{\rm
  thr}$, because one would have otherwise observed a similar
anisotropy pattern at an energy $E_{\rm thr}/Z$, with $Z$ the assumed
average charge of the cosmic rays at energies $>E_{\rm thr}$.

Consider a region of the sky, of angular size $\theta$ and solid angle
$\Delta\Omega=2\pi(1-\cos\theta)$, showing an excess of particles
above isotropic expectations above some energy threshold $E_{\rm
  thr}$. The most natural interpretation is to infer the existence of
one or more sources in this direction, the image of which is spread
over $\Delta\Omega$, because of smearing by $\theta$ in the
intervening magnetic fields, or because $\Delta\Omega$ subtends the
source distribution. For the sake of the argument, we assume that the
spectrum produced by the source consists of protons and of heavy
nuclei of charge $Z$. In all known source models, one expects that the
ratio of the elemental spectra $q_p(E)/q_Z(E)\,>\,1$ and more
generally $q_p(E)/q_Z(E)\,\gg\,1$ at a given energy $E\,\ll\,E_{\rm
  max}(p)$, where $E_{\rm max}(p)$ denotes the maximal energy at the
source for protons. For instance, the composition ratio of protons to
iron peak elements ($Z\geq 17$) in the Galactic cosmic ray spectrum,
which is roughly consistent with the solar abundance when compared at
a given energy per nucleon, implies a source composition at a given
energy $q_p:q_{Z\geq 17}\,\simeq\,1:0.06$ (as taken from the recent
ATIC-2 data, Panov et al. 2006). The only energy regime in which one
may obtain $q_p(E)/q_Z(E)\,\ll\,1$ is $E_{\rm
  max}(p)\,\ll\,E\,\ll\,E_{\rm max}(Z)$, with $E_{\rm max}(Z)$ the
maximal energy for the nuclei of charge $Z$ (see also further
below). We also assume here that the maximal energy scales as $Z$; as
argued further below, this is a conservative choice.  Now, let us ask
what would be seen if the composition above $E_{\rm thr}$ were
heavy. Consider the quantity
\begin{equation}
\Sigma_Z(>E_{\rm thr})\,\equiv\,\frac{\Delta N(>E_{\rm thr})}
{\sqrt{N_{\rm iso}(>E_{\rm thr})}}\ , \label{eq:sigma}
\end{equation}
which characterizes the signal-to-noise ratio of the anisotropy
signal: $\Delta N(>E_{\rm thr})$ represents the excess number of
events over isotropic expectations, i.e. the difference between the
total number of events observed above $E_{\rm thr}$ within
$\Delta\Omega$ and $N_{\rm iso}(>E_{\rm thr})$, the number of events
expected from the isotropic background. In the above model, $\Delta
N(>E_{\rm thr})$ corresponds on average to the number of events
produced by the sources inside $\Delta\Omega$. A detection of
anisotropy thus corresponds to $\Sigma_Z(>E_{\rm thr})\,\gg\,1$.

One can now compute a similar quantity, but corresponding to the
proton component at energies $>E_{\rm thr}/Z$, denoted
$\Sigma_p(>E_{\rm thr}/Z)$. These protons have a same magnetic
rigidity than the heavy component above $E_{\rm thr}$, therefore they
follow the same path in the Galactic and extra-galactic magnetic
fields, and consequently they are to produce a similar anisotropy
pattern in the sky, up to the composition ratio of the elemental
spectra and up to the level of background noise associated with the
isotropic component. Assuming that the elemental spectra at injection
have the same power-law form $q_i(E)\,\propto E^{-s}$ at
$E\,\ll\,E_{\rm max}(Z_i)$, and that the total number of events
observed scales as $E^{1-s_{\rm obs}}$, with $s_{\rm obs}$ the slope
of the observed all-sky differential spectrum, one finds
\begin{equation}
\Sigma_p(>E_{\rm thr}/Z)\,=\,\Sigma_Z(>E_{\rm thr})\frac{q_p(E_{\rm
    thr}/Z)}{q_Z(E_{\rm thr}/Z)}\,\alpha_{\rm loss}\,Z^{s-(s_{\rm
    obs}+1)/2}\ . \label{eq:sigmap}
\end{equation}
The fudge factor $\alpha_{\rm loss}\,\gtrsim\,1$ incorporates the
effect of energy losses; it is discussed in more detail in the
following.  At energies above the ankle, one has $s_{\rm
  obs}\,\simeq\,2.7$ (Abbasi et al. 2008, Abraham et
al. 2008a). Therefore, if the source emits a power-law with $s=2$, one
finds that the signal-to-noise ratio for detecting the anisotropy is
larger at $E_{\rm thr}/Z$ than at $E_{\rm thr}$ by a factor
$\approx\,\alpha_{\rm loss}\,Z^{0.2}q_p(E_{\rm thr}/Z)/q_Z(E_{\rm
  thr}/Z)$, generally expected to be significantly larger than unity!
Clearly, the softer the source spectrum, the larger the gain in
signal-to-noise ratio with decreasing $Z$.

The factor $\alpha_{\rm loss}$ accounts for the difference between the
propagated and injected spectra. Given that the energy loss distance
for protons of energy $E_{\rm thr}/Z$ is much larger than the energy
loss distance for heavy nuclei of energy $E_{\rm thr}$, if $E_{\rm
  thr}\,\gtrsim\,10^{19}\,$eV, one expects $\alpha_{\rm
  loss}>1$. Indeed, regarding the (primary) proton component, energy
losses can be safely neglected at $E_{\rm thr}/Z$ energies, since the
attenuation length is of order of a Gpc or more at EeV energies, which
is much larger than the depth up to which which anisotropies can be
produced, this latter being of order of the inhomogeneity scale of the
large scale structure $\sim 100\,$Mpc or of the distance to the
closest source, whichever is larger. Therefore, $\alpha_{\rm loss}$
approximately corresponds to the reciprocal of the attenuation factor
for heavy nuclei of charge $Z$ and energy $>E_{\rm thr}$ propagating
over the distance scale $l$ to the sources inside $\Delta\Omega$. The
attenuation factor is understood as the ratio of the flux above
$>E_{\rm thr}$ after propagation on a distance $l$ to the injected
flux above this energy.  For the case of iron primaries, this
attenuation factor is $0.7$ above $55\,$EeV at a distance of 10~Mpc,
becoming $0.4$ at a distance of 100~Mpc (counting in the propagated
flux all secondary nuclei with $Z>17$). For oxygen, the attenuation
factor is $0.5$ at 10~Mpc, becoming $0.1$ at 100~Mpc (counting in all
C, N and O secondary nuclei); see Bertone et al. (2002) for an
illustration of the relative fragility of ultra-high energy CNO nuclei
vs iron peak elements.

Actually, this enhancement factor $\alpha_{\rm loss}$ should be
further increased by the number of secondary protons with energy
$>E_{\rm thr}/Z$ produced by the photo-disintegration of the primary
heavy nuclei with energy $>E_{\rm thr}$. This number is not easy to
quantify but it can be significant. For instance, recent simulations
by Aloisio, Berezinsky \& Gazizov (2008) indicate that the signal
should be dominated by protons at energies $> 1-2\times 10^{19}\,$eV
even if the source composition were strongly enriched in iron nuclei
[assuming of course that $E_{\rm
    max}(p)\,>3\,\times10^{19}\,$eV]. Allard et al. (2008) also
conclude that heavy nuclei could be found in abundance at the detector
only if the composition were essentially dominated by Fe group nuclei
at energies beyond $10^{20}\,$eV. As discussed in this latter
reference, and above, the generic case that gives rise to a heavy
composition at ultra-high energies is when the proton maximal energy
at the source is smaller than the GZK cut-off. In the following, we
adopt a conservative point of view and neglect the effect of energy
losses, i.e. we set $\alpha_{\rm loss}=1$.

Let us now discuss the scaling of $E_{\rm max}$ with charge. The
maximal energy is given by the comparison of the acceleration
timescale (which depends on rigidity $E/Z$) with the minimum of the
escape timescale (which also depends on $E/Z$), the age of the source
or dynamical timescale (which do not depend on $E$, $Z$) and the
energy losses timescale (which depends on $E$, $Z$ and mass number
$A$). If the latter timescale does not provide the dominant limitation
to the acceleration mechanism, then the maximal energy scales as the
rigidity, as assumed above. If energy losses are dominant however, one
expects $E_{\rm max}(Z)\,\lesssim\,ZE_{\rm max}(p)$, as explained in
the following. Then the above argument would remain valid, as it only
requires that protons exist at energies $>E_{\rm thr}/Z$, i.e. that
$E_{\rm max}(p)\,\gtrsim\,E_{\rm thr}/Z$. Note that in the more
extreme case in which protons and heavy nuclei had similar maximal
energies, the composition at ultra-high energies $>E_{\rm thr}$ would
be dominated by protons, and therefore the anisotropy would be due to
those protons since they have a higher rigidity than the heavy nuclei
of a same energy.  For photo-hadronic interactions, the energy loss
timescales for protons and iron-peak nuclei are known to be
comparable, while that of intermediate elements is generally smaller;
in that case, one would expect $E_{\rm max}(p)\,\sim\,E_{\rm max}({\rm
  Fe})\,\gtrsim\,E_{\rm max}({Z\,\sim\,10})$. Regarding synchrotron
losses, the energy loss timescale scales as $(Z/A)^{-4}E^{-1}$,
therefore $E_{\rm max}\,\propto\,Z^{1/2}(Z/A)^{-2}$ if the
acceleration timescale $t_{\rm acc}\,\propto\,E/Z$ and synchrotron
losses dominate at the highest energies. In that case, the scaling
$E_{\rm max}(Z)/E_{\rm max}(p)\,\propto\, Z$ remains correct to within
a factor of order unity. The above conclusions agree with the recent
simulations of Allard \& Protheroe (2009).

Finally, accounting for a more complex chemical composition with more
than two species would not modify significantly the results shown in
Eq.~(\ref{eq:sigma}) as long as the anisotropy above $E_{\rm thr}$ is
indeed due to the species of charge $Z$. In practice, it will be
useful to test the composition above $E_{\rm thr}$ by checking the
anisotropies at various $E_{\rm thr}/Z_i$, with $Z_i\,=\,2,8,14,26$
representative of the most abundant elements.

In the following, we provide specific examples of the above test
applied to the most recent data of the PAO. We will show in particular
that the above argument remains valid when one considers the all-sky
anisotropy pattern due to an ensemble of close-by sources.

\section{Application to the Pierre Auger data}\label{sec:pao}
Abraham et al. (2007, 2008b) have reported a correlation of 20/27
arrival directions of the highest energy events ($E\,>\,5.7\times
10^{19}\,$eV) with nearby ($d\,<\,75\,$Mpc) active galactic
nuclei. However, the typical model of ultra-high energy cosmic ray
origin in AGN refers to strongly beamed Fanaroff-Riley II (FR~II)
sources with giant radio lobes (e.g. Rachen \& Biermann 1993), while
19 out of these 20 correlating AGN in the Pierre Auger dataset belong
to the Seyfert or LINER class, only one being a Fanaroff-Riley I
(FR~I) radio-galaxy (Cen~A). These correlating AGNs are thus more
likely to be tracers of the local large scale structure in which the
actual sources of ultra-high energy cosmic rays camouflage and indeed,
the arrival directions are compatible with a distribution of sources
following the large scale structure (Kashti \& Waxman 2008, Ghisellini
et al. 2008, Zaw, Farrar \& Greene 2009). The extended dataset up to
2009 of the PAO reveals a significantly weaker correlation, with only
26 out of 58 events above $55\,$EeV lying within $3.1^\circ$ of an AGN
located within 75\,Mpc (Hague et al. 2009). This level of correlation
is compatible with that expected for sources distributed as the large
scale structure (see discussion in Kotera \& Lemoine 2008 and Aublin
et al. 2009 for a detailed discussion of the correlation with
cosmological catalogs). It is important to note that the departure
from isotropy lies at the $1\%$ level. Therefore one should be careful
not to overinterpret this excess. In the following, we use these PAO
results as a concrete example for the test developed in the previous
section.

Interestingly, the PAO data up to 2007 has also shown evidence for
clustering of events in the region around Cen~A, with as many as $9$
events within $20^\circ$ (Gorbunov et al. 2008), as well as
correlation of $8$ out of 27 events with a sample of radio-galaxies
(Moskalenko et al.  2008; Nagar \& Matulich 2008).  This has triggered
a surge of interest in models of ultra-high energy cosmic ray origin in
Cen~A as well as forecast studies of neutrino and gamma-ray expected
signals from Cen~A (e.g. Cuoco \& Hannestad 2008, Gupta 2008, Halzen
\& O'Murchadha 2008, Holder et al. 2008, Kachelriess et
al. 2008). However, such a correlation does not necessarily imply that
Cen~A is the source of these cosmic rays. First of all, and as noted
in Abraham et al. (2008b), Hague et al. (2009), the evidence for
clustering is based on an a posteriori analysis, so that it is
difficult to quantify the level of significance. Even if future data
confirm this clustering, one must keep in mind that Cen~A lies in
front of one of the largest concentrations of matter in the local
Universe, at $\sim50\,$Mpc from us, so that the correlation could be
accidental. Finally, as we argue in Section~\ref{sec:cena}, the Cen~A
AGN/jet/lobe system is too weak to accelerate protons up to the
observed energies $>57\,$EeV.

Even more intriguing is the fact that the chemical composition
measured by the PAO becomes increasingly heavier above the ankle at
$4\,$EeV, with the last data points at $30-50\,$EeV lying close to
expectations for iron (Unger et al. 2007, Wahlberg et al. 2009). Such
a measurement is in obvious conflict with the most recent analysis of
HiRes data, which indicate a pure proton composition up to $30\,$EeV
(Belz 2009). As we argue in the following, it is also at odds with the
observed anisotropy pattern above $55\,$EeV, unless the composition
abruptly changes from heavy to light above $30-50\,$EeV.

\subsection{Anisotropy towards Cen~A}\label{sec:aniso-cena}
Let us first discuss the anisotropy toward Cen~A, using the most
recent PAO dataset, in particular, $12$ events located within
$18^\circ$ out of a total of $58$ events above $55\,$EeV (Hague et
al. 2009), assuming that the clustering is not a statistical
accident. In a first approximation, one can model the anisotropy
pattern as in the previous section, with one (or several clustered)
source(s) contributing a fraction $x$ of the total flux, the remaining
$1-x$ being accounted for by an isotropic background. We do not need
to assume that Cen~A itself is a source, but we center the anisotropic
signal on the location of Cen~A in order to match the observed
pattern.  Given that $2.7$ events are expected on average from this
region of the sky if the sources are distributed isotropically, we set
$x\,=\,0.15$ in the following.

We assume that the sources inject protons and iron nuclei with
powerlaw spectra $q_Z(E)\,\propto\,E^{-s}\exp\left\{-E/[ZE_{\rm
    max}(p)]\right\}$, producing an angular image of size
$\delta\theta$. As before, this angle could represent smearing due to
angular deflection, in which case $\delta\theta\,\propto\,(E/Z)^{-1}$,
or the angular size of the ensemble of sources. Although we assume
that all heavy nuclei are iron, the following could be easily
generalized to any kind of mixed composition without changing the
basic result. We also neglect the energy losses for the spectrum of
the anisotropic component, as before. This is motivated by the
proximity of matter in this direction of the sky and, as discussed
before, incorporating energy losses would not diminish the signal we
are to extract out of the data, while it would introduce other free
parameters. The remaining $1-x$ fraction of the flux is modelled with
an isotropic background, e.g. as produced by far away sources.

\begin{figure}
  \centering{
    \includegraphics[width=0.48\textwidth,clip=true]{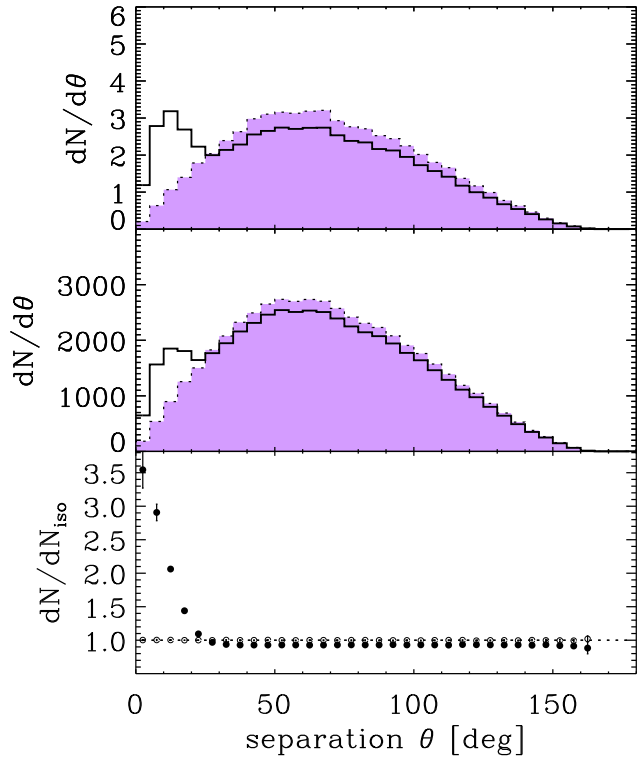}}
\caption{Top panel: differential histogram of the number of events
  above $55\,$EeV as a function of the angular separation to Cen~A, in
  solid line for the model described in the text (sources contributing
  15\% of the flux, forming an image of size $\delta\theta=10^\circ$,
  superimposed on an isotropic background) and expected signal from an
  isotropic population of sources (colored region). The flux is
  normalized to a total of 58 events above $55\,$EeV. Middle panel:
  same histogram but for $E>2.2\,$EeV, assuming that the high energy
  events are iron nuclei and that protons are injected with
  composition ratio $q_p(E):q_Z(E)=1:0.06$ at $E\,<\,3$EeV. The
  anisotropy signal is much stronger, as measured by the number of
  events in each bin. Bottom panel: in filled circles, the ratio of
  the differential histogram shown in the middle panel to the
  isotropic expectations ${\rm d}N_{\rm iso}/{\rm d}\theta$ with error
  bars estimated through the Monte Carlo; in empty circles, the same
  histogram but for a model assuming that the high energy events are
  protons only and $\delta\theta\propto (E/Z)^{-1}$. Only the latter
  model is compatible with an isotropic distribution of arrival
  directions at energies $>2.2\,$EeV.\label{fig:ptsa}}
\end{figure}

Let us first assume that the composition is dominated by iron at the
highest energies. We set $E_{\rm thr}=55\,$EeV, and
$\delta\theta=10^\circ$ above $E_{\rm thr}$. We also set $E_{\rm
  max}(p)=3\,$EeV, so that essentially no proton contribute to the
signal above $E_{\rm thr}$. In the top panel of Fig.~\ref{fig:ptsa} we
plot the histogram of the number of events as a function of the
angular distance to the source (solid line). The colored region
indicates the expectations for a purely isotropic signal. The
departure from isotropy is clear, although the uncertainty on the
signal in each bin is substantial given the small number of events
involved. Here the number of events within $18^\circ$ of Cen~A is
$\simeq9$ for the model (slightly lower than the observed 12 events,
yet within the uncertainty) and $2.7$ for the isotropic expectations
so that the inferred $\Sigma_{\rm Fe}(>55\,{\rm EeV})=\Delta
N/\sqrt{N_{\rm iso}}\,\simeq\,5.5$.

In the middle panel of Fig.~\ref{fig:ptsa}, we plot the same
histogram, but at an energy $E_{\rm thr}/26=2.2$EeV. In order to
simulate the anisotropy signal, we have proceeded as follows. We have
assumed that the total number of events increases (with decreasing
energy) according to the observed data (in particular the PAO spectrum
published in Yamamoto et al. 2007), and we have calculated the
contribution to this flux above $E_{\rm thr}/26$ for all the protons
injected by the source above this energy. We have assumed that the
ratio of protons to iron is 1:0.06 at a given energy well below
$E_{\rm max}$, corresponding to the composition ratio of proton to
iron peak elements ($Z\geq 17$) in the galactic cosmic ray spectrum
reconstructed at the source. The iron nuclei injected by the source at
energies $\gtrsim 2.2\,$EeV add up to the isotropic background and do
not contribute to the anisotropy signal. As made obvious in
Fig.~\ref{fig:ptsa}, the gain in signal-to-noise ratio is very large:
the number of events observed above $2.2\,$EeV within $18^\circ$ is
now 5150, for 2330 expected, so that $\Sigma_p(>2.2\,{\rm
  EeV})\,=\,\Delta N/\sqrt{N_{\rm iso}}\,\simeq\,100$.  This gain by a
factor $\simeq20$ can be recovered from Eq.~(\ref{eq:sigmap}).

Finally, the lower panel of Fig.~\ref{fig:ptsa} shows the ratio of the
differential histograms $dN/d\theta$ for the source proton
contribution plus isotropic background (filled circles) to the
isotropic background $dN_{\rm iso}/d\theta$ normalized to a same total
number of events, including the error bars evaluated through the Monte
Carlo simulation. The empty circles show the same quantity, but for a
model in which the source injects only protons, even up to the highest
energies, assuming that $\delta\theta$ represents magnetic deflection,
i.e. $\delta\theta\,\propto\,(E/Z)^{-1}$: we have set
$\delta\theta=10^\circ$ above $55\,$EeV in this model, so that the
anisotropy signal above $55\,$EeV is similar to that of the previous
iron+proton model shown in the top panel of Fig.~\ref{fig:ptsa}, but
the low energy signal is essentially isotropic, see the bottom panel.

The conclusions to be drawn are clear. If the clustering seen toward
Cen~A is real, the events responsible for this anisotropy pattern
toward Cen~A above $55\,$EeV are unlikely to be heavy nuclei,
otherwise PAO would have observed a much stronger anisotropy at
energies $\,<\,40\,$EeV (Abraham et al. 2007, 2008b).

\subsection{Correlation with the V\'eron-Cetty \& V\'eron (2006) catalog of AGN}\label{sec:aniso-agn}

\begin{figure}
  \centering{
    \includegraphics[width=0.48\textwidth,clip=true]{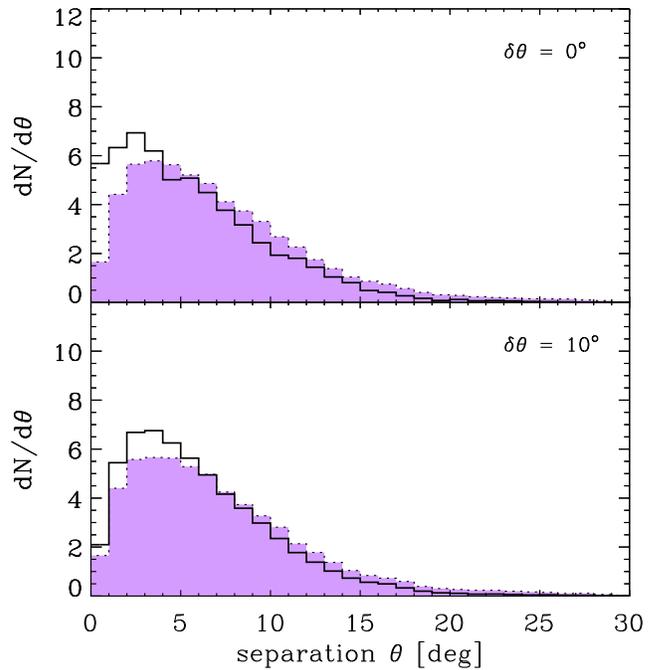}}
\caption{Differential histograms of the number of events above
  $55\,$EeV as a function of the angular separation to the closest
  AGN, drawn from the V\'eron-Cetty \& V\'eron (2006) catalog up to a
  distance of $75\,$Mpc. The solid line corresponds to the model in
  which 90\% of the flux above $55\,$EeV is produced by sources
  distributed as the large scale structure, modeled through the PSCz
  catalog of galaxies, smeared by a deflection angle $\delta\theta$,
  as indicated. The colored region indicates the expectation for
  purely isotropic arrival directions. The histograms are normalized
  to a total of 58 events, as observed by PAO.\label{fig:lssa}}
\end{figure}

In the present Section, we investigate how the anisotropy signal with
respect to nearby AGN, as reported in Hague et al. (2009) can be used
to discriminate the chemical composition at the highest energies. We
proceed in a similar way as before, but we now assume that the
spectrum is composed of two components: one that follows the local
large scale structure, contributing a fraction $x$ to the flux above
$55\,$EeV, and one isotropic component that makes up for the
remainder, $1-x$. As a proxy for the local large scale structure, we
use the PSCz catalog of galaxies (Saunders et al. 1991), as in
previous studies (Kashti \& Waxman 2008, Kotera \& Lemoine 2008). Out
of simplicity, we assume that the sources follow the PSCz distribution
without bias. However, we allow for the possibility of a deflection
angle $\delta\theta$ in intervening magnetic fields and we explore two
possible values $\delta\theta=0^\circ$ and $\delta\theta=10^\circ$
above $55\,$EeV. This choice is motivated by the resolution of the
PSCz catalog, approximately $7^\circ$, and by the possibility that
this signal is composed of heavy nuclei. In practice, we proceed
through Monte Carlo simulations as follows. We construct many mock
catalogs of angular sky distributions, each mock catalog being a
sample of $58$ events above $55\,$EeV (or the corresponding number at
any other energy). The arrival direction of each event in the sample
is drawn at random, according to a probability law reproducing the
PSCz distribution and PAO coverage if it belongs to the $x$
anisotropic fraction, or reproducing isotropic arrival directions and
PAO coverage if the event belongs to the remaining $1-x$ fraction of
isotropic background. In order to draw an event from the PSCz catalog,
we include the appropriate weight factors for the flux dilution with
inverse distance squared, for the PSCz exposure function and for the
PAO aperture. Then we randomly displace this arrival direction by
$\delta\theta$ in the plane of the sky. The energy losses are
accounted for by the depth of the catalog, which is limited to
$200~$Mpc.

We have set $x=0.90$, which allows to produce an anisotropy signal
close to the level observed: the PAO indeed records 24 events out of
58 events above $55\,$EeV located within $3^\circ$ of an AGN closer
than $75\,$Mpc, while the above model produces $19$ events within
$3^\circ$ if $\delta\theta=0^\circ$ and $14$ events if
$\delta\theta=10^\circ$. The model with 90\% of the flux above
$55\,$EeV coming from sources distributed as the large scale structure
with negligible deflection thus gives a number of correlating AGN
within $3^\circ$ that is comparable to that seen by PAO, to within the
uncertainty. A deflection angle $\delta\theta=10^\circ$ provides a
signal that is marginally too low, but given the simplifying
assumptions made above, this is not crucial. It certainly indicates
that larger deflection angles will probably result in too small
anisotropies. It is interesting to note that the anisotropy signal
peaks at a separation which is substantially smaller than
$\delta\theta$, i.e. $\sim4-5^\circ$ for $\delta\theta=10^\circ$. This
of course results from the PAO procedure, which selects the closest
AGN to the arrival direction and not to the source direction,
i.e. after the direction at the source has been modified by
$\delta\theta$.  The differential histograms corresponding to these
models and isotropic expectations are shown in Fig.~\ref{fig:lssa} as
a function of the angular separation to the closest AGN drawn from the
V\'eron-Cetty \& V\'eron (2006) catalog, with an arbitrary cut at
$75\,$Mpc, following the PAO analysis. These histograms are shown for
the two possible values of $\delta\theta$, as indicated. Here, we find
$\Sigma_{\rm Fe}(>55\,{\rm EeV})\,\simeq 2.1$ and $0.76$ for
$\delta\theta=0^\circ$ and $\delta\theta=10^\circ$ respectively. The
signal is thus marginal in terms of departure from isotropy.

\begin{figure}
  \centering{
    \includegraphics[width=0.48\textwidth,clip=true]{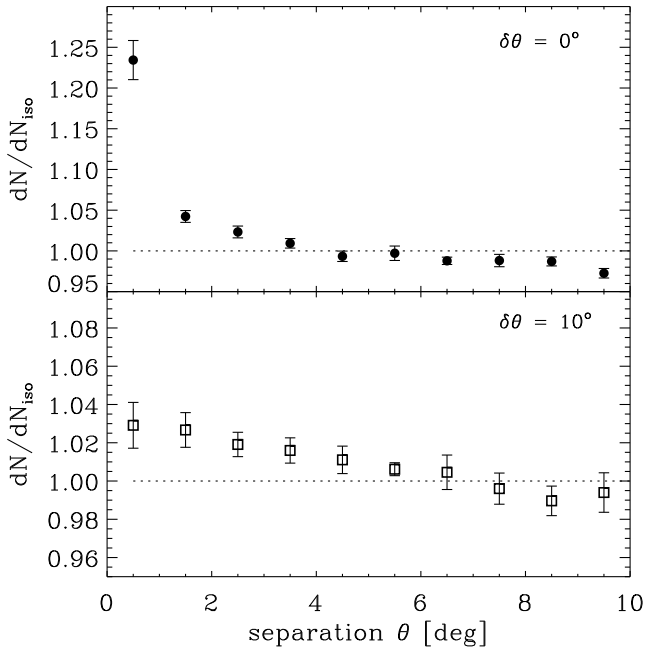}}
\caption{Ratio of the differential histograms for the models shown in
  Fig.~\ref{fig:lssa} to histograms for purely isotropic arrival
  directions, but computed from the proton component at energies
  $>2.2\,$EeV. The uncertainties measured through the Monte Carlo are
  indicated. Note the change of scale between the two panels
  corresponding to the two values of $\delta\theta$, as
  indicated.\label{fig:lssb}}
\end{figure}

We now plot the ratios of the differential histograms expected in
these two models at low energy $>2.2\,$eV to the histogram expected
for purely isotropic arrival directions, as before. We have assumed a
conservative composition ratio 1:0.06 between proton and iron as in
the previous section, and we have neglected energy losses on the
propagated spectra. The results are shown in Fig.~\ref{fig:lssb},
which clearly reveals a stronger anisotropy signal at these
energies. In terms of signal-to-noise, we find $\Sigma_p(>2.2\,{\rm
  EeV})\,\simeq\,6.2$ and $2.3$ respectively for
$\delta\theta=0^\circ$ and $\delta\theta=10^\circ$ ($\Delta N=623$ and
$235$ respectively, $N_{\rm iso}=10100$), which significantly exceed
the signal-to-noise ratios obtained above $55\,$EeV.  This generalizes
the argument presented in Section~\ref{sec:test-chem} for the case of
a more complex distribution of point sources on the sky, superimposed
on an isotropic background. It also shows that the present angular
data of the PAO disfavors a heavy composition at ultra-high
energies. Strictly speaking, this test demonstrates that the {\em
  anisotropic} part of the signal cannot be due to heavy nuclei {\em
  but} one cannot exclude that the isotropic background is made up of
heavy nuclei. However, our above model also suggests that a
significant fraction (90\% in Fig.~\ref{fig:lssa},\ref{fig:lssb}) of
the flux at $>55\,$EeV is produced by the anisotropic component if one
takes at face value the anisotropy results of the PAO. This suggests
that the bulk of ultra-high energy cosmic rays above $55\,$EeV are
protons.

\section{AGN as sources of ultra-high energy cosmic
  rays?}\label{sec:agn}

\subsection{AGN luminosity vs acceleration}\label{sec:lbound}

On general grounds, one can construct a lower bound to the magnetic
luminosity that a source must exhibit in order to be able to
accelerate particles up to $10^{20}\,$eV (Norman et al. 1995, Waxman
1995, Lyutikov \& Ouyed 2005, Waxman 2005).  Let us first briefly
describe the general, model independent argument given in Waxman
(1995).  Consider an astrophysical source driving a flow of magnetized
plasma, with characteristic magnetic field strength $B$ and velocity
$v$.  Imagine now a conducting wire encircling the source at radius
$R$.  The potential generated by the moving plasma is given by the
time derivative of the magnetic flux, and is therefore given by
$V=\beta B R$ where $\beta=v/c$. A proton which is allowed to be
accelerated by this potential drop would reach energy $E_p\sim\beta eB
R$. The situation is somewhat more complicated in the case of a
relativistic outflow, where $\Gamma\equiv(1-\beta^2)^{-1/2}\gg1$.  In
this case, the proton is allowed to be accelerated only over a
fraction of the radius $R$, comparable to $R/\Gamma$.  To see this,
one must realize that as the plasma expands, its magnetic field
decreases, so the time available for acceleration corresponds, say, to
the time of expansion from $R$ to $2R$.  In the observer frame this
time is $R/c$, while in the plasma rest frame it is $R/\Gamma
c$. Thus, a proton moving with the magnetized plasma can be
accelerated over a transverse distance $\sim R/\Gamma$.  This sets a
lower limit to the product of the magnetic field and source size,
which is required to allow acceleration to $E_p$, $BR>\Gamma
E_p/e\beta$.  This also sets a lower limit to the rate $L_B$ at which
energy is carried by the out flowing plasma, and which must be
provided by the source (Waxman 1995, 2005),
\begin{equation}\label{eq:L}
  L_B>\frac{\Gamma^2}{\beta}\left(\frac{E}{Ze}\right)^2c
  =10^{45.5}\frac{\Gamma^2}{\beta}\left(\frac{E/Z}{10^{20}\rm eV}\right)^2{\rm erg/s},
\end{equation}
where we have generalized the last equation to particles of charge
$Z$.  

Let us consider next the application of the general argument given
above to the commonly assumed scenario, where particle acceleration
within the outflowing plasma is achieved via diffusive acceleration.
In this case, one writes the acceleration timescale $t_{\rm
  acc}\,\equiv\,{\cal A}t_{\rm L}$ as a multiple ${\cal A}$ of the
Larmor time, and one compares it to the dynamical timescale available
for acceleration. Let us discuss this bound in some detail. For the
sake of definiteness, we assume an outflow of Lorentz factor $\Gamma$
and opening angle $\Theta$ (the isotropic case is recovered in the
limit $\Theta^2\rightarrow 2$).  The quantity ${\cal A}$ depends on
the diffusion coefficient of the particle, hence on its energy and the
general properties of the magnetic field of the accelerating
region. In acceleration mechanisms which involve scattering against
magnetic inhomogeneities (such as Fermi I or Fermi II), $t_{\rm
  acc}\,\propto t_{\rm s}$, where $t_{\rm s}=3\kappa/c$ is the
scattering timescale and $\kappa$ the diffusion coefficient.  The
parallel diffusion coefficient can be written as $\kappa/(r_{\rm
  L}c)\,\simeq\,\eta^{-1}\left(r_{\rm L}/\lambda_B\right)^\alpha$ (see
for instance Casse, Lemoine \& Pelletier 2002), with $\eta<1$ the
fraction of turbulent magnetic energy density (in units of total
magnetic energy density) and $\lambda_B$ the coherence length. The
exponent $\alpha$ depends on the spectral index $\beta$ of the (one
dimensional) turbulence power spectrum: $\alpha=\beta-1$ for $r_{\rm
  L}<\lambda_B$, $\alpha=1$ for $r_{\rm L}>\lambda_B$. From this, one
easily verifies that $\kappa/(r_{\rm L}c)>1$ is general, meaning that
the so-called Bohm regime $\kappa=r_{\rm L}c$ is a limiting regime.
In non-relativistic Fermi shock acceleration ($\Gamma_{\rm
  sh}\beta_{\rm sh}\,\la\,1$ or $\beta_{\rm sh}\la 0.7$), $t_{\rm
  acc}\,\simeq\,3\kappa/\beta_{\rm sh}^2$, with $\beta_{\rm sh}$ the
shock velocity in the comoving frame and $\kappa$ refers to quantities
measured in the upstream frame. Hence ${\cal A}\,\simeq 3\beta_{\rm
  sh}^{-2}\kappa/(r_{\rm L}c)\,\gg\,1$ . One recovers a similar result
for non-relativistic Fermi II acceleration, with ${\cal A}\,\sim\,
\beta_{\rm A}^{-2}\kappa/(r_{\rm L}c)\,\gg\,1$, $\beta_{\rm A}<1$
denoting the Alfv\'en speed. If acceleration takes place in a
(non-relativistic) shear flow with velocity gradient $\Delta u/\Delta
x$, the timescale $t_{\rm acc}\,\sim\, \Delta x^2/(\Delta u^2 t_{\rm
  s})$ (Rieger, Bosch-Ramon \& Duffy 2007). The deconfinement limit
corresponds to $t_{\rm s}\,\sim\, \Delta x/c$, in which case the
limiting acceleration timescale becomes comparable to $t_{\rm
  s}c^2/\Delta u^2$, as above. Note that the acceleration timescale is
larger (hence acceleration is less efficient) when $t_{\rm s} \,<\,
\Delta x/c$. In the ultra-relativistic shock limit $\Gamma_{\rm
  sh}\beta_{\rm sh}\,\gg\,10$, one expects ${\cal A}\,\simeq\,
\Gamma_{\rm sh}^{-2}r_{\rm L}/\lambda_B$ (up to a factor of order
unity, see Pelletier, Lemoine \& Marcowith 2009), with the condition
$r_{\rm L}/\lambda_B\,\gg\,\Gamma_{\rm sh}$. Since ${\cal A}$
increases with $r_{\rm L}$, this process is in general not efficient.
Finally, in the mildly relativistic regime $1\,\la\,\Gamma_{\rm
  sh}\beta_{\rm sh}\,\la\,10$, one expects $t_{\rm acc}\,\sim\, t_{\rm
  s}$, hence ${\cal A}\,\simeq\,\kappa/(r_{\rm L}c)$, whose minimum
value is of order one provided the Bohm regime diffusion applies and
the magnetic field is nearly fully turbulent.

To summarize, ${\cal A}\,\gg\,1$ in general, and ${\cal
  A}\,\approx\,1$ corresponds to a maximally efficient acceleration
process.  The constraint $t_{\rm acc}\,\leq\,t_{\rm dyn}$, with
$t_{\rm dyn}=R/\beta\Gamma c$ the dynamical timescale at radius $R$
(in the lab frame), now gives:
\begin{equation}
L_B\,\geq\, 0.65\times 10^{45}\, \Theta^2\Gamma^2{\cal
  A}^2\beta^3Z^{-2}E_{20}^2\, {\rm erg/s}\ ,
\end{equation}
with $E_{20}$ the observed energy in units of $10^{20}\,$eV.  The
presence of $\beta^3$ does not mean that in the limit
$\beta\rightarrow 0$, this lower bound vanishes, since ${\cal
  A}^2\propto \beta_{\rm sh}^{-4}$ more than compensates for this
term.  Similarly, the above bound does not vanish in the limit
$\Theta\rightarrow 0$, since lateral escape losses become prominent
when $\Theta\,<\,\Gamma^{-1}$. The corresponding timescale $t_{\rm
  esc}\,\simeq\, (\Theta R)^2/2\kappa$, hence $L_B\,\geq\,1.2\times
10^{45}\,{\cal A}\beta (\kappa/r_{\rm L}c)Z^{-2}E_{20}^2 \,{\rm
  erg/s}$.  In the case of AGN jets, one finds in general
$\Theta\Gamma\sim1$, hence both limits are comparable. According to
the above discussion, the most optimistic values for the parameters
entering these equations are $\beta\sim 1$, ${\cal A}\,\sim\,1$ and
$\kappa/(r_{\rm L}c)\,\sim\,1$.  Therefore $10^{45}Z^{-2}\,$erg/s can
be considered as a firm lower bound on the source magnetic luminosity.

Since most of the AGN seen to correlate within $3^\circ$ of the PAO
events are Seyfert galaxies, with bolometric luminosities well below
$10^{45}\,$ergs/s, the AGN that they harbor cannot be the source of
light particles at $10^{20}\,$eV. We also note that Zaw, Farrar \&
Greene (2009) have shown that one third of this sample is actually
made of star forming galaxies with little or no AGN activity. Hence
the correlation with stricto-sensu AGN objects is actually weaker than
reported. 

Actually even FR~I radio-galaxies, TeV blazars and BL Lac objects do
not possess the required power to accelerate $Z\sim1$ particles up to
$10^{20}\,$eV, since their magnetic luminosities are of order
$L_B\sim10^{42}-10^{44}\,$ergs/s (Celotti \& Ghisellini 2008). Such
objects might possibly accelerate heavy nuclei up to energies around
the GZK cut-off but, as discussed above, the anisotropy pattern at
high and especially low energies would then appear inconsistent with
the PAO results. Further anisotropy studies in both energy ranges
would certainly further constrain their possible contribution.

The compilation of Celotti \& Ghisellini (2008) has been done in the
frame of leptonic synchrotron and inverse Compton emission. In
hadronic blazar models, the magnetic field in the blazar zone is
typically one order of magnitude larger than in leptonic models, so
that these objects might fulfill the magnetic luminosity bound to
accelerate protons up to $10^{20}\,$eV. However, in this case,
acceleration would occur in the blazar zone hence the emission should
be beamed forward. In order to escape further expansion losses in the
magnetized jets, the accelerated protons would have to be converted
into neutrons, which would decay back to protons on a distance scale
$\sim0.9E_{20}\,$Mpc, i.e. outside the jet. One should therefore
observe a correlation of the arrival directions with blazars, not with
radio-galaxies seen offside (Rachen 2008).  However the Pierre Auger
Observatory reports no correlation with blazars (Harari et al. 2007),
so that this possibility is equally at odds with present observational
results.

Keeping in line with the modelling of Celotti \& Ghisellini (2008),
only flat spectrum radio quasars (the likely equivalent to the most
powerful FR~II sources) seem capable of producing jets with
$L_B\,>\,10^{45}\,$ergs/s. But, considering the highest energy PAO
event ($E=1.48\pm0.27\pm0.32\times 10^{20}\,$eV), one finds that the
smallest angular separation of this event to the FR~II sources located
within 130~Mpc, as compiled by Massaglia (2007) is already $28^\circ$
(for NGC~4261). The next objects are 3C~296 ($38^\circ$ away) and
PKS1343-60 ($41^\circ$ away). The closest blazar (classified as BL Lac
in the V\'eron \& V\'eron-Cetty catalog) located closer than
$150\,$Mpc lies  $115^\circ$ away from this highest energy event
(TEX~0554+534).

Recently, it has been proposed that transients in active galactic
nuclei could power up the engine up to the luminosities required,
thereby evading the above constraints (see for instance Farrar \&
Gruzinov 2009, Dermer et al. 2008). However, as demonstrated by Waxman
\& Loeb (2009), such transients should produce counterpart flares in
X-rays through the concomittant acceleration of electrons, which
should have been picked up by existing surveys. Their non-detection
strongly argues against such flaring scenarios.

\subsection{The particular case of Cen~A}\label{sec:cena}
Given its proximity and the apparent clustering of PAO events in this
direction of the sky, Cen~A has received a lot of attention. In
particular, several authors have suggested that this AGN could host a
site of acceleration to ultra-high energies (Cavallo 1978; Romero et
al.  1996; Farrar \& Piran 2000; Gorbunov et al.  2008; Hardcastle et
al. 2009, O'Sullivan, Reville \& Taylor 2009). Let us therefore
discuss it in some more detail.

Centaurus~A is classified as a FR~I / misaligned blazar (see Israel
1998 for a review). Its bolometric luminosity $L_{\rm bol}\,\sim\,
10^{43}\,$erg/s and its jet kinetic power is inferred to be $L_{\rm
  jet}\,\simeq\,2\times 10^{43}\,$erg/s. Therefore, this source should
not be able to accelerate light particles to $10^{20}\,$eV. The
detailed modelling of the spectral energy distribution of its nucleus
by Chiaberge, Capetti \& Celotti (2001) provides further information
on the emission zone: $R\,\sim\, 10^{16}\,$cm, $B\,\sim\,0.5\,$G and
$\Gamma\,\leq\,3-5$. This corresponds to a magnetic luminosity
$L_B\,=\,\frac{1}{4} R^2B^2\Gamma^2\beta c\,\sim\,2\times
10^{42}\,$ergs/s, a value that seems quite reasonable in view of the
jet kinetic power and of the bolometric luminosity, and which agrees
relatively well with estimates of the magnetic field in the inner jet
at larger distances from the core, when assuming $BR\,\sim\,$constant.

The papers which have discussed the issue of acceleration to
ultra-high energy in some detail are Romero et al. (1996), which has
argued that acceleration can take place in the X-ray knots of the
inner jet through diffuse shock acceleration, Hardcastle et al. (2009)
and O'Sullivan, Reville \& Taylor (2009), which have discussed the
stochastic acceleration of protons in the giant lobes of Cen~A and
Gorbunov et al. (2008) relying on the model of Neronov et al. (2007)
of particle acceleration in a black hole magnetosphere.

Romero et al. (1996) uses the inferred values of $B\,\simeq\,
10^{-5}\,$G (upstream field), $R\,\simeq\,1.8\,$kpc, $\beta_{\rm
  sh}\,\simeq\,0.3$ and $\eta\,\simeq\,0.4$. However, since the Larmor
radius $r_{\rm L}\,\simeq\,11\,{\rm
  kpc}\,E_{20}Z^{-1}B_{-5}^{-1}\,>\,R$ (with $E_{20}=E/10^{20}\,$eV,
$B_{-5}=B/10^{-5}\,$G), even the Hillas criterion (Hillas 1984) is not
satisfied. Balancing the acceleration timescale with the escape
timescale $R/c$, one derives a maximal energy
$E\,\la\,10^{18}\,Z\beta_{0.3}^2B_{-5}R_{1.8}\,$eV
($R_{1.8}\,\equiv\,R/1.8\,{\rm kpc}$, $\beta_{0.3}\,\equiv\,\beta_{\rm
  sh}/0.3$). As discussed above, this maximal energy lies in the
correct ballpark for accelerating nuclei to ultra-high
energies. However the anisotropy signal would then appear incompatible
with the PAO data; in particular, a strong anisotropy signal at EeV
energies should have been detected.

Hardcastle et al. (2009) have argued that protons could be accelerated
up to $\sim 5\times 10^{19}\,$eV through stochastic acceleration in
the magnetized turbulence of the giant lobes of Cen~A. However, one
can show that their estimate of the acceleration timescale is far too
optimistic because, when applied to the electrons, it produces a
maximal energy well in excess of the inferred $E_{{\rm
    max},e}\,\sim\,1-4\times10^{11}\,$eV. In detail, using the
estimate of Hardcastle et al. (2009) for the acceleration timescale,
$t_{\rm acc}\,\simeq\,3.5\times10^6\,{\rm yrs}\,E_{20}B_{-6}^{-1}$,
and their Bohm scaling $t_{\rm acc}\,\propto E$, one infers the
maximal energy for electrons by comparing $t_{\rm acc}$ to the
synchrotron and inverse Compton energy loss timescales, leading to
$E_{{\rm max},e}\,\sim\,2\times10^{16}\,B_{-6}^{-1/2}\,$eV.  The above
maximal electron energy is five orders of magnitude larger than the
inferred maximal energy, hence the above acceleration timescale is far
too optimistic and, consequently, the maximal proton energy must be
much lower than $10^{20}\,$eV. This conclusion agrees with those of
O'Sullivan, Reville \& Taylor (2009); see also Casse, Lemoine \&
Pelletier (2002) for a similar discussion regarding the acceleration
in the lobes of powerful radio-galaxies.

Finally, Neronov et al. (2007) have argued that in a magnetic field of
strength $B=10^4\,$G threading a maximally rotating black hole of mass
$M_{\rm bh}=10^8\,M_\odot$, heavy nuclei could be accelerated to
energies of order $10^{20}\,$eV by exploiting the potential drop
$\Phi\,\simeq\, 10^{20}B_4M_{8}\,$V. However, this maximal energy is
quite optimistic because it ignores radiative energy losses, which are
highly efficient in the nuclei of powerful AGN (Norman et al. 1995,
Henri et al. 1999). In any case, the Poynting flux that is required,
$L_{\rm B}\,\sim\, 10^{45}\,$erg/s far exceeds any other luminosity
measured elsewhere in the galaxy, and in particular the estimate
obtained from multi-band spectral modelling of the nucleus emission
(Chiaberge, Capetti \& Celotti 2001).

\section{Discussion}\label{sec:disc}

\subsection{Summary}
Let us start by summarizing the results obtained so far. We have
discussed how one can can test the chemical composition of ultra-high
rays on the sky by comparing the anisotropy signals at various
energies. Our main result is to show that if anisotropies are observed
above some energy $E_{\rm thr}$ and the composition is assumed to be
heavy at that energy (nuclei of charge $Z$), one should observe at
energies $>E_{\rm thr}/Z$ a substantially stronger anisotropy signal,
which is associated with the proton component emitted by the sources
that are responsible for the anisotropy pattern. These conclusions do
not depend at all on the modelling of the intervening magnetic fields
and they are robust with respect to the parameters characterizing the
sources.

We have then applied this test to the most recent data of the PAO. We
find that, if these data are taken at face value (and if the PAO
arrival directions around the ankle appear isotropic), the events that
are responsible for the anisotropy signal reported toward Cen~A and
toward the nearby AGN should not be heavy nuclei. Interestingly, this
result appears to be at odds with the current PAO results on the
chemical composition at the highest energies. One cannot exclude at
the present time that the observed anisotropies are statistical
accidents; only the acquisition of a larger set of data will tell. One
cannot exclude either that the composition switches abruptly from
heavy to light at $\sim30\,$EeV. Finally, one cannot exclude that a
systematic bias affects the composition measurements of the PAO.  In
any case, the present discussion indicates that, if both the
anisotropy and the heavy composition are confirmed by future data,
some possibly important information is to be extracted out of this
apparent contradiction.

The theoretical and experimental uncertainties in the extrapolation of
the $p-p$ cross-section to center-of-mass energies
$\sqrt{s}\,\gtrsim\,100\,$TeV are one of the possible sources of
biases in shower reconstruction. Recently, it has been argued that, if
this cross-section were underestimated by some $40-60\,$\% at these
energies, one might reconcile the existing $X_{\rm max}$ measurements
with a pure proton composition at energies above the ankle (Ulrich et
al. 2009a, 2009b; see however Wibig 2009). In this respect, and in the
light of the above discussion, the apparent anisotropy towards Cen~A
offers a very valuable opportunity to test this possibility. It
suffices, in principle, to test the anisotropy pattern against a
future (independent) data set at various energies by comparing the
number of events detected within a predetermined solid angle in a
predetermined direction to the expected number for isotropic arrival
directions; at the same time, one needs to perform a dedicated
composition measurement for cosmic rays from this particular region of
the sky, in order to avoid as much as possible the contamination due
to the isotropic contribution to the all-sky flux. Then, one should
apply the test discussed above.

\subsection{The interpretation of the PAO data}

The natural question that follows is what can be inferred about the
sources of ultra-high energy cosmic rays if, following the anisotropy
results of PAO and the above discussion, one assumes that most
ultra-high energy cosmic rays are protons. Let us now explore this
line of reasoning.
 
First of all, one concludes that the existing data of the PAO disfavor
the AGN model of ultra-high energy cosmic ray origin. Indeed, we have
discussed in Section~\ref{sec:agn} a robust lower bound to the
magnetic luminosity that a source must possess in order to be able to
accelerate particles of charge $Z$ up to $10^{20}\,$eV,
$L_B\,\gtrsim\,\,10^{45}\,Z^{-2}\,$erg/s. This bound allows to rule
out the local radio quiet AGN as sources of $10^{20}\,$eV protons.  We
have also argued that, while FR~I radio-galaxies do not seem to
possess the required luminosity to accelerate protons up to
$10^{20}\,$eV, FR~II radio-galaxies and blazars do not correlate with
the highest energy events. Although radio-loud galaxies could possibly
accelerate heavy nuclei up to energies close to the GZK cut-off, as we
have discussed they would produce an anisotropy pattern predominantly
at $\sim\,$EeV energies rather than at ultra-high energies, in
conflict with the PAO results.

We have discussed the particular case of Cen~A in some detail, as this
source has recently received a lot of attention due to the observed
clustering of events around it. We have provided a critical discussion
of acceleration in this object and argued that its AGN/jet/lobe system
cannot produce protons at the energies required, in good agreement
with the general expectations from the above luminosity bound. That
being said, the excess clustering in this particular direction, if
real, remains to be interpreted.

First, if sources are distributed according to the large scale
structure, one expects a certain number of events to coincide
accidentally with the direction of Cen~A, because of the location of
the Centaurus and Shapley superclusters in this direction. Using the
PSCz catalog of galaxies as a proxy for the distribution of the
sources, with a depth of $200\,$Mpc, and accounting for the PAO
exposure, one expects $0.80$ events above $55\,$EeV within $6^\circ$
of Cen~A nucleus (and $0.3$ for isotropic arrival directions) for
$2-3$ observed, or $4.4$ events within $18^\circ$ (and $2.7$ for
isotropic arrival directions) for $12$ observed. Note that the large
scale structure is poorly sampled in the proximity of Cen~A due to its
low galactic latitude. Nevertheless, the above numbers already
indicate that the contamination from the surrounding large scale
structure in the direction to Cen~A is substantial. In particular, at
the present level of statistics, one cannot claim observing a
significant departure from a model in which the sources are
distributed as the large scale structure. It should also be recalled here
that the PAO analysis of clustering around Cen~A is an a posteriori
one, hence one may not therefore assign a reliable significance to the
detection of the anisotropy.

Moreover, if sources are gamma-ray bursts or magnetars, one should
include in the flux prediction the contribution of such sources
located in the Cen~A host galaxy itself. It is well known that the
probability of detecting ultra-high energy cosmic rays from nearby
gamma-ray bursts is extremely small unless the arrival times of these
ultra-high energy cosmic rays are sufficiently dispersed 
(Waxman 1995, Waxman \& Miralda-Escud\'e 1996). Consider for instance
gamma-ray bursts with rate of occurence $\dot N_{\rm GRB}$ in Cen~A;
these can be seen if either of their jets points into a solid angle
$2\pi(1-\cos\delta\theta)$ where $\delta\theta$ is the typical
deflection acquired by the cosmic ray en route to the detector, giving
a detection probability $P\,\sim\, \dot N_{\rm
  GRB}\sigma_t\delta\theta^2/2\,\sim\,10^{-4}\dot N_{{\rm
    GRB},-5}E_{70}^{-4}B_{-8}^4\lambda_{100{\rm kpc}}^2$, with $\dot
N_{\rm GRB}\,=\,10^{-5}\dot N_{{\rm GRB},-5}\,{\rm yr}^{-1}$;
$\sigma_t\,=\,6\times10^3\,{\rm yr}\,B_{-8}^2\lambda_{100{\rm
    kpc}}E_{70}^{-2}$ represents the spread of arrival times and
$\delta\theta=3^\circ B_{-8}\lambda_{100{\rm kpc}}^{1/2}E_{70}^{-1}$
the angular deflection for particles of energy $E=70\,E_{70}\,$EeV
traveling from Cen~A to the detector through a magnetic field of
strength $10^{-8}B_{-8}\,$G and coherence length $100\,\lambda_{\rm
  100\,kpc}\,$kpc (Waxman \& Miralda-Escud\'e 1996). The above
probability suggests that one should not detect events from gamma-ray
bursts emitted directly toward the observed. However, if the jet of
the gamma-ray burst hits one of the lobes of Cen~A, the particles can
be redistributed in all directions, in particular towards the
detector. The magnetic field inside the lobe, $B\sim1\,\mu$G is indeed
sufficient to impart a deflection of order unity on the lobe distance
scale $\sim 100\,$kpc to protons of $70\,$EeV. In this case, the
expected time delay and time dispersion are of the order of the travel
time to the lobes, $L_{\rm lobes}/c\,\sim\,3\times 10^5\,$yr, so that
the expected number of gamma-ray bursts that one can see at any time,
through rescattering of the particles on the lobes of Cen~A is:
\begin{equation}
\langle N_{\rm GRB}\rangle\,\simeq\,\dot N_{\rm GRB}\frac{L_{\rm lobes}}{c}
\frac{\Delta\Omega_{\rm lobes}}{2\pi}\,\sim\,{\cal O}(1)\dot N_{{\rm GRB},-5}
\end{equation}
for an apparent solid angle of the lobes $\Delta\Omega_{\rm lobes}$ as
viewed from the host galaxy of Cen~A, which is close to $4\pi$.

The last gamma-ray burst(s) of Cen~A may then contribute to the flux
above the GZK cut-off up to:
\begin{equation}
j_{\rm Cen\,A}\,\approx\,0.9\times10^{-41}\,{\rm eV}^{-1}{\rm cm}^{-2}{\rm
  s}^{-1}\,
\epsilon_{51}E_{70}^{-2}f_\epsilon^{-1}\sigma_{t,5.5}^{-1}\langle
N_{\rm GRB}\rangle\ ,
\end{equation}
with $\sigma_{t,5.5}=\sigma_{t}/3\times10^5\,$yrs,
$f_\epsilon=\ln(E_{\rm max}/E_{\rm min})\,\sim\,1-10$ for an injection
spectrum with $s=2.0$, and $\epsilon=10^{51}\epsilon_{51}\,$ergs the
energy injected by one gamma-ray burst in ultra-high energy cosmic
rays. This flux should be spread over the angular scale of the lobes,
$\sim 10^\circ$. At $70\,$EeV, PAO records a diffuse flux $j_{\rm
  PAO}(70\,{\rm EeV})\,\simeq\,3\times10^{-40}\,{\rm eV}^{-1}{\rm
  cm}^{-2}{\rm s}^{-1}{\rm sr}^{-1}$. After correcting for the solid
angle of the Cen~A image and the PAO aperture towards Cen~A, one finds
that the last gamma-ray bursts in Cen~A should contribute up to
$2-25\,$\% of the observed PAO flux, well within the range of the
observed anisotropy.

We also note that this contribution from sources inside Cen~A would
improve the overall correlation of all the PAO data above $55\,$EeV
with sources distributed as the local large scale structure. Finally,
we remark that the above effect of rescattering on the lobes provides
a clear example of the possible biases discussed in Kotera \& Lemoine
(2008), namely that the PAO could be preferentially observing the last
scattering centers (in the present case, the lobes of Cen~A) rather
than the source itself (the gamma-ray bursts located in the core of
Cen~A).

\acknowledgments 












\begin{thebibliography}{}
\bibitem[]{2004ApJ...610L..73A} Abbasi, R.~U., et al. (The HiRes
  Collaboration), 2004, \apjl, 610, L73
\bibitem[]{2005ApJ...623..164A} Abbasi, R.~U., et al. (The HiRes
  Collaboration), 2005, \apj, 623, 164
\bibitem[]{2008PhRvL.100j1101A} Abbasi, R.~U., et al. (The HiRes
  Collaboration), 2008, Physical Review Letters, 100, 101101
\bibitem[]{2007Sci...318..938T} Abraham, J. et al. (The Pierre Auger
  Collaboration), 2007, Science, 318, 938
\bibitem[]{2008PhRvL.101f1101A} Abraham, J., et al. (The Pierre Auger
  Collaboration), 2008a, Physical Review Letters, 101, 061101 
\bibitem[]{2008APh....29..188T} Abraham, J. et al. (The Pierre Auger
  Collaboration), 2008b, Astroparticle Physics, 29, 188
\bibitem[]{2008JCAP...10..033A} Allard, D., Busca, N.~G., Decerprit,
  G., Olinto, A.~V., \& Parizot, E.\ 2008, Journal of Cosmology and
  Astro-Particle Physics, 10, 33
\bibitem[]{2009arXiv0902.4538A} Allard, D., \& Protheroe, R.~J.\ 2009,
  arXiv:0902.4538
\bibitem[]{2008arXiv0803.2494A} Aloisio, R., Berezinsky, V., \&
  Gazizov, A.\ 2008, arXiv:0803.2494
\bibitem[]{2003ApJ...589..871A} Arons, J.\ 2003, \apj, 589, 871
\bibitem[]{JB09} Belz, J. (The HiRes Collaboration), {\it proc.
  CRIS-2008} (Malfa, Italy, Sept. 2008)
\bibitem[]{Aea09} Aublin, J. et al. (The Pierre Auger
  Collaboration), 2009, arXiv:0906.2347, ICRC 2009
\bibitem[]{2002PhRvD..66j3003B} Bertone, G., Isola, C., Lemoine, M.,
  \& Sigl, G.\ 2002, \prd, 66, 103003
\bibitem[]{2002PhRvD..65b3002C} Casse, F., Lemoine, M., \& Pelletier,
  G.\ 2002, \prd, 65, 023002
\bibitem[]{1978A&A....65..415C} Cavallo, G.\ 1978, \aap, 65, 415
\bibitem[]{CG08} Celotti, A., Ghisellini, G., 2008, MNRAS, 385, 283
\bibitem[]{2001MNRAS.324L..33C} Chiaberge, M., Capetti, A., \&
  Celotti, A.\ 2001, \mnras, 324, L33
\bibitem[]{2008PhRvD..78b3007C} Cuoco, A., \& Hannestad, S.\ 2008, \prd,
      78, 023007
\bibitem[]{2008arXiv0811.1160D} Dermer, C.~D., Razzaque, S., Finke,
  J.~D., \& Atoyan, A., 2008, arXiv:0811.1160
\bibitem[]{EFS1} Evans, N. W., Ferrer, F., Sarkar, S., 2003, PRD 69, 3501.
\bibitem[]{EFS2} Evans, N. W., Ferrer, F., Sarkar, S., 2004, PRD 69, 8302.
\bibitem[]{2000astro.ph.10370F} Farrar, G.~R., \& Piran, T.\ 2000,
  arXiv:astro-ph/0010370
\bibitem[]{2006ApJ...642L..89F} Farrar, G.~R., Berlind, A.~A., \&
  Hogg, D.~W.\ 2006, \apjl, 642, L89
\bibitem[]{2009ApJ...693..329F} Farrar, G.~R., \& Gruzinov, A., 2009,
  \apj, 693, 329
\bibitem[]{2004APh....21..359F} Finley, C.~B., \& Westerhoff,
  S.\ 2004, Astroparticle Physics, 21, 359
\bibitem[]{2008arXiv0806.2393G} Ghisellini, G., Ghirlanda, G.,
  Tavecchio, F., Fraternali, F., \& Pareschi, G.\ 2008,
  arXiv:0806.2393
\bibitem[]{GTT1} Gorbunov, D. S., Tinyakov, P. G., Tkachev, I. I.,
Troitsky, S. V., 2002, ApJ 577, L93.
\bibitem[]{GTTT} Gorbunov, D. S., Tinyakov, P. G., Tkachev, I. I.,
Troitsky, S. V., 2004, {\tt astro-ph/0406654}.
\bibitem[]{2008arXiv0804.1088G} Gorbunov, D.~S., Tinyakov, P.~G.,
  Tkachev, I.~I., \& Troitsky, S.~V.\ 2008, arXiv:0804.1088
\bibitem[]{gzk} Greisen, K., 1966, PRL 16, 748.
\bibitem[]{2008JCAP...06..022G} Gupta, N.\ 2008, Journal of
Cosmology and Astro-Particle Physics, 6, 22
\bibitem[]{Hea09} Hague, J. D. et al. (The Pierre Auger
  Collaboration), arXiv:0906.2347, ICRC 2009
\bibitem[]{2008arXiv0802.0887H} Halzen, F., \& O'Murchadha, A.\ 2008,
      arXiv:0802.0887
\bibitem[]{2007arXiv0706.1715H} Harari, D., et al. (The Pierre Auger
  Collaboration), 2007, arXiv:0706.1715
\bibitem[]{2008arXiv0808.1593H} Hardcastle, M.~J., 
Cheung, C.~C., Feain, I.~J., \& Stawarz, L.\ 2009, MNRAS, 393, 1041
\bibitem[]{1999APh....11..347H} Henri, G., Pelletier, G., Petrucci,
  P.~O., \& Renaud, N.\ 1999, Astroparticle Physics, 11, 347
\bibitem[]{1984ARA&A..22..425H} Hillas, A.~M.\ 1984, \araa, 22, 425
\bibitem[]{2008arXiv0810.0471H} Holder, J., \& for the VERITAS
      Collaboration 2008, arXiv:0810.0471
\bibitem[]{1998A&ARv...8..237I} Israel, F.~P.\ 1998, \aapr, 8, 237
\bibitem[]{2008arXiv0805.2608K} Kachelriess, M., Ostapchenko, S., \&
Tomas, R.\ 2008, arXiv:0805.2608
\bibitem[]{2008JCAP...05..006K} Kashti, T., \& Waxman, E.\ 2008,
  Journal of Cosmology and Astro-Particle Physics, 5, 6
\bibitem[]{2008PhRvD..77l3003K} Kotera, K., \& Lemoine, M.\ 2008,
  \prd, 77, 123003
\bibitem[]{2005astro.ph..7620L} Lyutikov, M., \& Ouyed, R.\ 2005,
  arXiv:astro-ph/0507620
\bibitem[]{Massaglia} Massaglia, S., 2007, Nuc. Phys. B
  Proc. Supp. 168, 302
\bibitem[]{1995ApJ...449L..37M} Milgrom, M., \& Usov, V.\ 1995, \apjl,
  449, L37
\bibitem[]{2008arXiv0805.1260M} Moskalenko, I.~V., Stawarz, L.,
  Porter, T.~A., \& Cheung, C.~C.\ 2008, arXiv:0805.1260
\bibitem[]{2008arXiv0806.3220N} Nagar, N.~M., \& Matulich, J.\ 2008,
  arXiv:0806.3220
\bibitem[]{2007arXiv0712.1737N} Neronov, A., Semikoz, D., \& Tkachev,
  I.\ 2007, arXiv:0712.1737
\bibitem[]{1995ApJ...454...60N} Norman, C.~A., Melrose, D.~B., \&
  Achterberg, A.\ 1995, \apj, 454, 60
\bibitem[]{OSRT09} O'Sullivan, S., Reville, B., \& Taylor, A. M.,
  2009, arXiv:0903.1259
\bibitem[]{PLM08} Pelletier, G., Lemoine, M., \& Marcowith, A., 2008,
  arXiv:0707.3459
\bibitem[Panov et al.(2006)]{2006astro.ph.12377P} Panov, A.~D., et
  al. (ATIC-2 Collaboration) 2006, arXiv:astro-ph/0612377
\bibitem[]{1993A&A...272..161R} Rachen, J.~P., \& Biermann,
  P.~L.\ 1993, \aap, 272, 161
\bibitem[]{} Rachen, J., 2008, in Proceedings of ``Rencontres de Blois
  2008 - Challenges in particle astrophysics'', Blois (France), May
  18-23 2008, ed. J. Dumarchez
\bibitem[]{2007Ap&SS.309..119R} Rieger, F.~M., Bosch-Ramon, V., \&
  Duffy, P.\ 2007, \apss, 309, 119
\bibitem[]{1996APh.....5..279R} Romero, G.~E., Combi, J.~A., Perez
  Bergliaffa, S.~E., \& Anchordoqui, L.~A.\ 1996, Astroparticle
  Physics, 5, 279
\bibitem[]{1991Natur.349...32S} Saunders, W., Frenk, C.,
  Rowan-Robinson, M., Lawrence, A., \& Efstathiou, G.\ 1991, \nat,
  349, 32
\bibitem[]{S95} Stanev, T., Biermann, P. L., LLoyd-Evans, J., Rachen,
J. P., Watson, A. A., 1995, PRL 75, 3056.
\bibitem[]{1999ApJ...522..225T} Takeda, M., et al.\ 
1999, \apj, 522, 225 
\bibitem[]{TT01} Tinyakov, P. G., Tkachev, I. I., 2001, Pisma
Zh. Eksp. Teor. Fiz. 74, 3 [JETP Lett. 74, 1].
\bibitem[]{tinyakov} Tinyakov, P. G., Tkachev, I. I., 2002, Astropart. Phys. 18, 165.
\bibitem[]{TT2} Tinyakov, P. G., Tkachev, I. I., 2004, PRD 69, 8301.
\bibitem[]{U00} Uchihori, Y., Nagano, M., Takeda, M., Teshima, M.,
LLoyd-Evans, J., Watson, A. A., 2000, Astropart. Phys. 13, 151.
\bibitem[]{2009arXiv0906.0418U} Ulrich, R., Engel, R., M{\"u}ller, S.,
  Pierog, T., Sch{\"u}ssler, F., \& Unger, M.\ 2009a, arXiv:0906.0418
\bibitem[]{2009arXiv0906.3075U} Ulrich, R., Engel, R., 
M{\"u}ller, S., Sch{\"u}ssler, F., \& Unger, M.\ 2009b, arXiv:0906.3075 
\bibitem[]{U97} Unger, M. et al. (The Pierre Auger Collaboration), ICRC-2007, arXiv:0706.1495
\bibitem[]{2006A&A...455..773V} V{\'e}ron-Cetty, M.-P., \& V{\'e}ron,
  P.\ 2006, \aap, 455, 773
\bibitem[]{1995ApJ...453..883V} Vietri, M.\ 1995, \apj, 453, 883
\bibitem[]{Wea09} Wahlberg, H. et al. (The Pierre Auger
  Collaboration), 2009, arXiv:0906.2319, ICRC 2009
\bibitem[]{W95} Waxman, E. 1995, Phys. Rev. Lett., 75, 386.
\bibitem[]{1996ApJ...472L..89W} Waxman, E., \& Miralda-Escud\'e,
  J.\ 1996, \apjl, 472, L89
\bibitem[]{W05} Waxman, E., 2005, Phys. Scripta, T121, 147
\bibitem[]{2008arXiv0809.3788W} Waxman, E., \& Loeb, A., 2008,
  arXiv:0809.3788
\bibitem[]{2008arXiv0810.5281W} Wibig, T.\ 2008, arXiv:0810.5281
\bibitem[]{2009PhRvD..79i4008W} Wibig, T.\ 2009, \prd, 79, 
094008
\bibitem[]{Yea07} Yamamoto, T. et al. (The Pierre Auger
  Collaboration), ICRC 2007
\bibitem[]{ZK66} Zatsepin, G.~T., Kuzmin, V.~A., 1966, Pis'ma
Zh. Eksp. Teor. Fiz. 4, 114 [JETP. Lett. 4, 78].
\bibitem[]{2009ApJ...696.1218Z} Zaw, I., Farrar, G.~R., 
\& Greene, J.~E.\ 2009, \apj, 696, 1218
\end{thebibliography}
\end{document}